\documentclass{aa}

\usepackage{graphicx,epsfig,epsf}
\usepackage{longtable}
\usepackage{amssymb}

\begin{document}
\renewcommand{\topfraction}{0.85}
\renewcommand{\bottomfraction}{0.7}
\renewcommand{\textfraction}{0.15}
\renewcommand{\floatpagefraction}{0.66}

\newcommand{\hess}{H.E.S.S.}
\newcommand{\msh}{MSH 15-5\textit{2}}

\title{Discovery of extended VHE gamma-ray emission from the asymmetric pulsar wind nebula in \msh\ with \hess\ }

\titlerunning{Discovery of VHE $\gamma$-ray emission from \msh\ with \hess}

\author{F. Aharonian\inst{1}
 \and A.G.~Akhperjanian \inst{2}
 \and K.-M.~Aye \inst{3}
 \and A.R.~Bazer-Bachi \inst{4}
 \and M.~Beilicke \inst{5}
 \and W.~Benbow \inst{1}
 \and D.~Berge \inst{1}
 \and P.~Berghaus \inst{6} \thanks{Universit\'e Libre de 
 Bruxelles, Facult\'e des Sciences, Campus de la Plaine, CP230, Boulevard
 du Triomphe, 1050 Bruxelles, Belgium}
 \and K.~Bernl\"ohr \inst{1,7}
 \and C.~Boisson \inst{8}
 \and O.~Bolz \inst{1}
 \and I.~Braun \inst{1}
 \and F.~Breitling \inst{7}
 \and A.M.~Brown \inst{3}
 \and J.~Bussons Gordo \inst{9}
 \and P.M.~Chadwick \inst{3}
 \and L.-M.~Chounet \inst{10}
 \and R.~Cornils \inst{5}
 \and L.~Costamante \inst{1,20}
 \and B.~Degrange \inst{10}
 \and A.~Djannati-Ata\"i \inst{6}
 \and L.O'C.~Drury \inst{11}
 \and G.~Dubus \inst{10}
 \and D.~Emmanoulopoulos \inst{12}
 \and P.~Espigat \inst{6}
 \and F.~Feinstein \inst{9}
 \and P.~Fleury \inst{10}
 \and G.~Fontaine \inst{10}
 \and Y.~Fuchs \inst{13}
 \and S.~Funk \inst{1}
 \and Y.A.~Gallant \inst{9}
 \and B.~Giebels \inst{10}
 \and S.~Gillessen \inst{1}
 \and J.F.~Glicenstein \inst{14}
 \and P.~Goret \inst{14}
 \and C.~Hadjichristidis \inst{3}
 \and M.~Hauser \inst{12}
 \and G.~Heinzelmann \inst{5}
 \and G.~Henri \inst{13}
 \and G.~Hermann \inst{1}
 \and J.A.~Hinton \inst{1}
 \and W.~Hofmann \inst{1}
 \and M.~Holleran \inst{15}
 \and D.~Horns \inst{1}
 \and O.C.~de~Jager \inst{15}
 \and B.~Kh\'elifi \inst{1}
 \and Nu.~Komin \inst{7}
 \and A.~Konopelko \inst{1,7}
 \and I.J.~Latham \inst{3}
 \and R.~Le Gallou \inst{3}
 \and A.~Lemi\`ere \inst{6}
 \and M.~Lemoine-Goumard \inst{10}
 \and N.~Leroy \inst{10}
 \and T.~Lohse \inst{7}
 \and O.~Martineau-Huynh \inst{16}
 \and A.~Marcowith \inst{4}
 \and C.~Masterson \inst{1,20}
 \and T.J.L.~McComb \inst{3}
 \and M.~de~Naurois \inst{16}
 \and S.J.~Nolan \inst{3}
 \and A.~Noutsos \inst{3}
 \and K.J.~Orford \inst{3}
 \and J.L.~Osborne \inst{3}
 \and M.~Ouchrif \inst{16,20}
 \and M.~Panter \inst{1}
 \and G.~Pelletier \inst{13}
 \and S.~Pita \inst{6}
 \and G.~P\"uhlhofer \inst{1,12}
 \and M.~Punch \inst{6}
 \and B.C.~Raubenheimer \inst{15}
 \and M.~Raue \inst{5}
 \and J.~Raux \inst{16}
 \and S.M.~Rayner \inst{3}
 \and I.~Redondo \inst{10,20}\thanks{now at Department of Physics and
Astronomy, Univ. of Sheffield, The Hicks Building,
Hounsfield Road, Sheffield S3 7RH, U.K.}
 \and A.~Reimer \inst{17}
 \and O.~Reimer \inst{17}
 \and J.~Ripken \inst{5}
 \and L.~Rob \inst{18}
 \and L.~Rolland \inst{16}
 \and G.~Rowell \inst{1}
 \and V.~Sahakian \inst{2}
 \and L.~Saug\'e \inst{13}
 \and S.~Schlenker \inst{7}
 \and R.~Schlickeiser \inst{17}
 \and C.~Schuster \inst{17}
 \and U.~Schwanke \inst{7}
 \and M.~Siewert \inst{17}
 \and H.~Sol \inst{8}
 \and R.~Steenkamp \inst{19}
 \and C.~Stegmann \inst{7}
 \and J.-P.~Tavernet \inst{16}
 \and R.~Terrier \inst{6}
 \and C.G.~Th\'eoret \inst{6}
 \and M.~Tluczykont \inst{10,20}
 \and G.~Vasileiadis \inst{9}
 \and C.~Venter \inst{15}
 \and P.~Vincent \inst{16}
 \and H.J.~V\"olk \inst{1}
 \and S.J.~Wagner \inst{12}}

\institute{
Max-Planck-Institut f\"ur Kernphysik, Heidelberg, Germany
\and
 Yerevan Physics Institute, Armenia
\and
University of Durham, Department of Physics, U.K.
\and
Centre d'Etude Spatiale des Rayonnements, CNRS/UPS, Toulouse, France
\and
Universit\"at Hamburg, Institut f\"ur Experimentalphysik, Germany
\and
APC, Paris, France 
\thanks{UMR 7164 (CNRS, Universit\'e Paris VII, CEA, Observatoire de Paris)}
\and
Institut f\"ur Physik, Humboldt-Universit\"at zu Berlin, Germany
\and
LUTH, UMR 8102 du CNRS, Observatoire de Paris, Section de Meudon, France
\and
Groupe d'Astroparticules de Montpellier, IN2P3/CNRS, Universit\'e Montpellier II, France 
\and
Laboratoire Leprince-Ringuet, IN2P3/CNRS, Ecole Polytechnique, Palaiseau, France
\and
Dublin Institute for Advanced Studies, Ireland
\and
Landessternwarte, K\"onigstuhl, Heidelberg, Germany
\and
Laboratoire d'Astrophysique de Grenoble, INSU/CNRS, Universit\'e Joseph Fourier, France 
\and
DAPNIA/DSM/CEA, CE Saclay, Gif-sur-Yvette, France
\and
Unit for Space Physics, North-West University, Potchefstroom, South Africa
\and
Laboratoire de Physique Nucl\'eaire et de Hautes Energies, IN2P3/CNRS, Universit\'es Paris VI \& VII, France
\and
Institut f\"ur Theoretische Physik, Lehrstuhl IV: Weltraum und Astrophysik,
    Ruhr-Universit\"at Bochum, Germany
\and
Institute of Particle and Nuclear Physics, Charles University, Prague, Czech Republic
\and
University of Namibia, Windhoek, Namibia
\and
European Associated Laboratory for Gamma-Ray Astronomy, jointly supported by CNRS and MPG
}

\offprints{Bruno.Khelifi@mpi-hd.mpg.de}

\date{Accepted by A\&A}

\abstract{

The Supernova Remnant \msh\ has been observed in very high energy (VHE) $\gamma$-rays using the \hess\ 4-telescope array located
in Namibia. A $\gamma$-ray signal is detected at the $25$ sigma level during an exposure of 22.1 hours live time. The image
reveals an elliptically shaped emission region around the pulsar PSR\,B1509--58, with semi-major axis $\sim$$6'$ in the NW-SE
direction and semi-minor axis $\sim$$2'$. This morphology coincides with the diffuse pulsar wind nebula as observed at X-ray
energies by ROSAT. The overall energy spectrum from $280$\,GeV up to $40$\,TeV can be fitted by a power law with photon index
$\Gamma$$=$$2.27$$\pm$$0.03_{\textrm {\scriptsize \,stat}}$$\pm$$0.20_{\textrm{\scriptsize \,syst}}$. The detected emission can be
plausibly explained by inverse Compton scattering of accelerated relativistic electrons with soft photons.

 }

\maketitle

\section{Introduction}

The Supernova Remnant (SNR) \msh\ (G\,320.4--1.2), first observed as an extended non-thermal radio source by Mills et al.\,
(\cite{radio1}), is a complex object with an unusual morphology. Radio observations by Caswell et al.\ (\cite{radio2}) reveal a
roughly circular SNR $\sim$$30'$ in diameter with a bright feature in the NW rim and a fainter one in the SE. The $10'$ diameter
NW source (G\,320.4--1.0) coincides with the $H_{\alpha}$ nebula RCW\,89.

Einstein X-ray observations of \msh\ by Seward and Harnden (\cite{einstein}) led to the discovery of the 150-ms pulsar
PSR\,B1509--58 located within the SNR shell, surrounded by a diffuse extended non-thermal component. The existence of a pulsar
wind nebula (PWN) was later confirmed by ROSAT (\cite{trussoni}; Brazier \& Becker \cite{brazier}) and ASCA (\cite{tamura}) as an
extended emission region surrounding the pulsar with a power-law photon index of $\sim$$2.0$. The PWN morphology is clearly
visible in the ROSAT PSPC data (\cite{trussoni}) which show an elongated structure roughly centered on the pulsar with two arms
extending several arcminutes along the NW and SE directions. These features were more recently confirmed by detailed Chandra
observations (Gaensler et al.\ \cite{chandra}). 

Gaensler et al.\ (\cite{atca}), comparing X-ray and radio observations, concluded that the radio SNR \msh\ and the pulsar
PSR\,B1509--58 are parts of a single system interacting via a pair of opposed collimated outflows (i.e. the PWN). The distance
to the object and the pulsar spindown age were estimated to $5.2$$\pm$$1.4$\,kpc and 1700 years, respectively.

The PWN hard X-ray spectrum was measured by BeppoSAX (\cite{mineo}) and is fitted by a power law with photon index of
$2.08$$\pm$$0.01$ in the 1--200\,keV range. Interpreted as synchrotron emission from electrons within the PWN, these results
suggest the presence of electrons accelerated up to energies of tens of TeV, possibly leading to VHE $\gamma$-ray emission through
an inverse Compton (IC) process as suggested by du Plessis et al.\ (\cite{duplessis}). Early observations by the CANGAROO
experiment (\cite{sako}) yielded a marginal $\gamma$-ray signal above 1.9\,TeV corresponding to $\sim$$10$\% of the flux of the
Crab Nebula. 

\section{\hess\ Observations and Results}

\begin{figure*}[!t]
\vspace{-0.3cm}
\centering
\includegraphics[width=13.0cm]{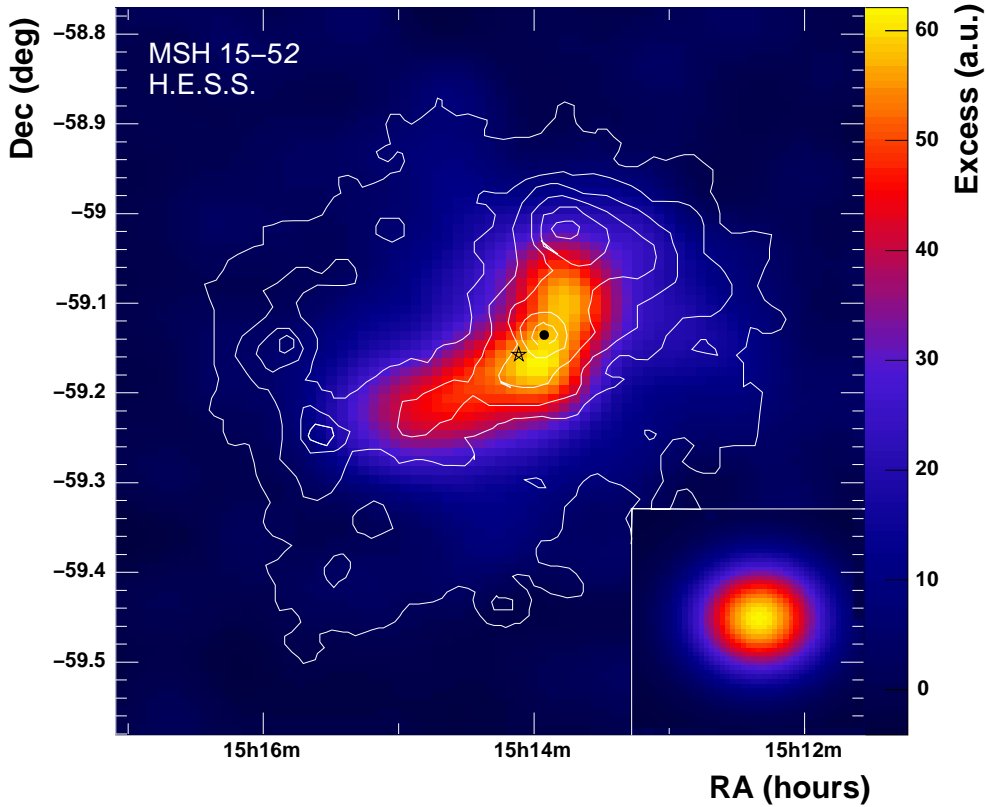}
\vspace{-0.2cm}
\caption{Smoothed excess map from \msh\ in arbitrary units (a.u.). The map is smoothed with a Gaussian of $\sigma$$=$$0.04^\circ$ 
and only events with image sizes above 400 p.e. are used in order to improve the \hess\ angular resolution. The white contour
lines denote the X-ray (0.6--2.1\,keV) count rate measured by ROSAT (\cite{trussoni}). The black point and black star lie at the
pulsar position and at the excess centroid, respectively. The right-bottom inset shows the simulated PSF smoothed identically.}
\vspace{-0.4cm}
\label{fig_ex}
\end{figure*}

\msh\ was observed using the atmospheric Cherenkov technique with the full 4-telescope \hess\ ({\em High Energy Stereoscopic
System}) array (\cite{hinton}). The observations were made between March and June 2004 in {\it wobble mode}, for which the target
position is offset by $0.5^\circ$ with respect to the centre of the field of view. Only data with good atmospheric conditions were
selected yielding a total live time of 22.1\,hours. The selected data were taken at a mean zenith angle of $37^\circ$. A standard
analysis is used for the selection of $\gamma$-ray candidates, for the stereoscopic direction reconstruction of showers, and for
spectral analysis (\cite{2155}), resulting in an average energy threshold of $\sim$$280$\,GeV at this zenith angle. For each sky
bin ($0.6'$$\times$$0.6'$), the background level is estimated from events falling in a ring centered on this test position with a
mean radius of $1^\circ$ and with an area 7 times that of the cell.

An excess with a significance of $\sim$$25\,\sigma$ (using the likelihood method of Li \& Ma \cite{lima}) is detected within the
region of radius $0.14^\circ$ centered at the pulsar position with the standard point source analysis. Figure~\ref{fig_ex} shows
the excess map of $\gamma$-ray candidates with image sizes greater than 400 photoelectrons (p.e.). The latter cut is meant to
reduce the angular resolution to $\le$$0.07^\circ$ (70\% containment radius) and raises the energy threshold to $\sim$$900$\,GeV.
The map is convolved with a Gaussian of $\sigma$$=$$0.04^\circ$ in order to smooth out statistical fluctuations. An extended
emission is seen along the jet axis of the pulsar (\cite{trussoni}) in the NW-SE direction. There is no evidence for VHE
$\gamma$-ray emission from the shell of SNR G\,320.4--1.2, including the NW part (RCW\,89) from which a combination of thermal and
non-thermal emission was seen in X-rays by ASCA (\cite{tamura}) and Chandra (Gaensler et al.\ \cite{chandra}).

\begin{figure}[!t]
\centering
\includegraphics[width=9cm]{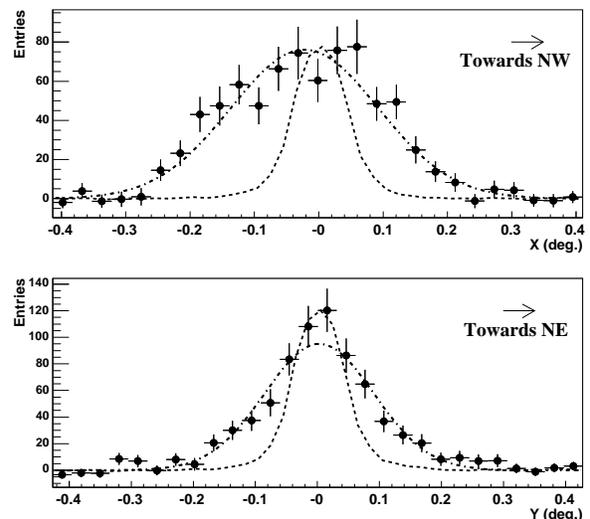}
\vspace{-0.8cm}
\caption{Projection of the unsmoothed excess map along the major (top) and minor (bottom) axes relative to the pulsar position. 
The dot-dashed lines are the best-fit Gaussians to these distributions, and the dashed lines show the distribution for a
Monte-Carlo point source excess at the pulsar position.}
\label{fig_proj}
\vspace{-0.6cm}
\end{figure}

The unsmoothed excess map (with image sizes above 400 p.e.) is fitted by a two-dimensional Gaussian convolved with the point
spread function (PSF) in order to extract the intrinsic dimensions. The best fit position of the Gaussian centroid is
(15h14m7s$\pm$21s, $-59^\circ9'27''$$\pm$$11''$) which is displaced by more than $3$$\,\sigma$ from the pulsar position
(15h13m55.6s, $-59^\circ8'8.9''$) taking into account the $20''$ systematic error on the pointing in each direction. The fitted
direction of the major axis is $41^\circ$$\pm$$13^\circ$ with respect to the RA axis, and is compatible with the X-ray direction
($60^\circ$$\pm$$5^\circ$) observed by Chandra. The intrinsic standard deviations along the major and minor axis are
$6.4'$$\pm$$0.7'$ and $2.3'$$\pm$$0.5'$, respectively. These dimensions are comparable to those of the diffuse PWN as observed by
ROSAT, $\sim$$10'$$\times$$6'$ (\cite{trussoni}). Figure~\ref{fig_proj} shows the projection along the major and minor axis of the
count map relative to the pulsar position and illustrates the morphological features described above. The transverse profile is
approximately symmetric. The longitudinal profile, in contrast, appears to extend more in the SE than in the NW direction relative
to the pulsar position. Given the spatial coincidence and the similar morphology of the \hess\ picture and the X-ray data, we
identify this new VHE $\gamma$-ray source with the pulsar wind nebula in \msh.

The total $\gamma$-ray excess within a circle centered on the Gaussian centroid with a radius of $0.3^\circ$, which encompasses
the entire VHE $\gamma$-ray emission, is $3481$$\pm$$129$ events after standard cuts. Fig.~\ref{fig_spec} shows the
reconstructed spectrum from these events. The data are consistent with a power law ($dN/dE$$\propto$$E^{-\Gamma}$) up to
$\sim$$40$\,TeV with a photon index of $\Gamma$$=$$2.27$$\pm$$0.03_{\textrm{\scriptsize \,stat}}$$\pm$$0.20_{\textrm{\scriptsize
\,syst}}$ and a differential flux at $1$\,TeV of $(5.7$$\pm$$0.2_{\textrm{\scriptsize \,stat}}$$\pm$$1.4_{\textrm{\scriptsize
\,syst}}) \times 10^{-12} \,\textrm{TeV}^{-1}\,\textrm{cm}^{-2}\,\textrm{s}^{-1}$. The $\chi^2$/d.o.f. of the fit is 13.3/12.
The corresponding integral flux above $280$\,GeV represents $\sim$$15$\% of the Crab Nebula flux above the same threshold. The
flux distribution, as measured on a run by run basis, is compatible with steady emission (the $\chi^2$/d.o.f. is 47.0/51).

\section{Discussion and Conclusions}
\label{discu}

\begin{figure}[t]
\vspace{-0.3cm}
\centering
\includegraphics[width=8cm]{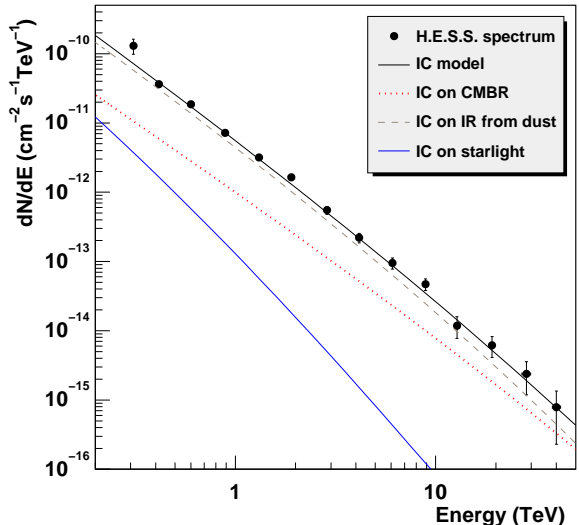}
\vspace{-0.5cm}
\caption{Reconstructed VHE $\gamma$-ray spectrum of the whole nebula. The lines are the best model fit of the one zone
model described in section \ref{discu}.}
\label{fig_spec}
\vspace{-0.49cm}
\end{figure}

The \hess\ observations of \msh\ provide the first image of an extended PWN in the VHE range. The emission region is clearly
extended along the jet axis. If this VHE $\gamma$-ray emission arises from inverse Compton (IC) scattering of target photons by
electrons, the latter should emit synchrotron radiation below
1\,keV. Thus, the morphology similar to that seen in soft X-rays
by ROSAT suggests a leptonic origin of the \hess\ signal. Assuming the target photons are uniformly distributed, IC emission
directly reflects the distribution of high-energy electrons, unlike synchrotron emission which also reflects spatial variations
of the magnetic field.

The lack of VHE $\gamma$-ray emission in the NE-SW direction shows that the elongated morphology of the PWN in X-rays is not due
to a lower magnetic field strength, $B$, further away from the jet axis. However the reason for the apparent lack of high-energy
electrons in the equatorial direction is not clear. If the spin axis is aligned with the magnetic axis of the pulsar, as suggested
by Brazier \& Becker (\cite{brazier}), Sulkanen \& Lovelace (\cite{sulkanen}) show that most of the energy, angular momentum and
electric current are carried within the jet, which may enhance the accelerated electron density wihtin jets. The VHE $\gamma$-ray
emission asymmetry along the jet axis relative to the pulsar may be interpreted as Doppler boosting within the two opposite jets
(\cite{pelling}). The two jets may also be confined differently by the surrouding medium, as suggested by the appearance of
RCW\,89 to the NW.

The total power, $L$, radiated by \msh\ in the energy band $0.3$--$40$\,TeV is $\sim$$1.0 \times 10^{35}$\,erg\,s$^{-1}$
(here we adopt a distance of 5\,kpc) compared to $\sim$$8.0 \times 10^{34}$\,erg\,s$^{-1}$ for the Crab Nebula (assuming a
distance of 2\,kpc). Thus, the efficiency of VHE $\gamma$-ray production relative to the spin-down rate ($L/\dot{E}$) of
PSR\,B1509--58 ($\sim$$0.6$\%) is much larger than that of the Crab pulsar ($\sim$$0.016$\%). Such high efficiency has been
proposed previously (\cite{pwn}).

A simple one-zone IC model (\cite{khelifi}) can be used to reproduce the VHE spectrum of the whole nebula. Therein, a population
of accelerated electrons with a power-law energy spectrum is assumed. These particles lose their energy by synchroton emission
in a magnetic field, assumed to be uniform within the PWN, and by IC scattering on seed photons. The electron spectrum and the
magnetic field strength are adjusted such that the computed synchrotron and IC spectra match the X-ray and VHE $\gamma$-ray
observations. The energy spectrum of the whole nebula measured by BeppoSAX (\cite{mineo}) is used for this fit. By using only
the cosmic microwave background radiation (CMBR) as seed photons, it is not possible to reproduce the VHE spectrum. Thus,
infrared (IR) photons from dust and starlight are added, using a recent parametrisation of the interstellar radiation field
(ISRF) used for the GALPROP code (\cite{GALPROP}). The energy density of the dust and starlight components of the ISRF are
however kept free to account for possible local variations. 

The best fit model is represented in Figure \ref{fig_spec} in the VHE band with the contribution of the different seed photons
to the IC spectrum.  The fitted spectral index of radiating electrons is $2.9$. The fitted mean magnetic field strength is
$\sim$$17\,\mu$G, which is about a factor of two higher than the lower limit estimated previously (Gaensler et al.\
\cite{chandra}). The best fit energy density of the dust component is about $2.3$\,eV\,cm$^{-3}$ which is more than twice the
nominal value of the GALPROP ISRF. This apparent enhancement of dust density may be due to the material swept up by the SNR
G\,320.4--1.2 (du Plessis et al.\ \cite{duplessis}). As the energy density of the starlight component is unconstrained, it is
fixed to the GALPROP value ($\sim$$1.4$\,eV\,cm$^{-3}$). 

Although this spectral model is compatible with all available measurements, the derived value of $B$, when combined with the
pulsar spin-down age of 1700 years, suggests the existence of a cooling break in the electron spectrum of $\sim$$24$\,TeV. This
corresponds to a photon energy above $\sim$$3$\,TeV in the IC spectrum, depending on the energy of the seed photons, which
should be detectable. Nonetheless, if the age of the system is closer to $20\,000$ years as suggested by Gvaramadze (\cite{age}),
the above spectral model would be  self-consistent. It should also be noted that there is noticeable correlation between the
fitted $B$ value and the photon energy density from dust. Thus, if this local density was higher, the corresponding $B$ value
would increase and thus the cooling break energy in the IC spectrum would decrease.

In summary, this detection represents the first morphological identification of an extended pulsar wind nebula  in the VHE
$\gamma$-ray band. The \hess\ observations provide more direct information on accelerated particles than soft X-ray data, due to
the uncertainties in the estimation of the magnetic field and its spatial variations.

\begin{acknowledgements}

The support of the Namibian authorities and of the University of Namibia in facilitating the construction and
operation of \hess\ is gratefully acknowledged, as is the support by the German Ministry for Education and
Research (BMBF), the Max-Planck-Society, the French Ministry for Research, the CNRS-IN2P3 and the Astroparticle
Interdisciplinary Programme of the CNRS, the U.K. Particle Physics and Astronomy Research Council (PPARC), the
IPNP of the Charles University, the South African Department of Science and Technology and National Research
Foundation, and by the University of Namibia. We appreciate the excellent work of the technical support staff in
Berlin, Durham, Hamburg, Heidelberg, Palaiseau, Paris, Saclay, and in Namibia in the construction and operation
of the equipment. We would also like to thank T. Mineo for provision of the BeppoSAX data in numerical form.

\end{acknowledgements}

\end{document}